# Entitymetrics: Measuring the impact of entities

Ying Ding, Min Song, Jia Han, Qi Yu, Erjia Yan, Lili Lin, Tamy Chambers

Ying Ding, Department of Information and Library Science, Indiana University, Bloomington, IN, USA. dingying@indiana.edu
Min Song, Department of Library and Information Science, Yonsei University. min.song@yonsei.ac.kr
Jia Han. Department of Information and Library Science, Indiana University, Bloomington, IN, USA. hanjia@imail.iu.edu.
Qi Yu, Department of Information Management, Shanxi Medical University, Taiyuan, China. yuqi351@yahoo.com.cn.
Erjia Yan, College of Information Science and Technology, Drexel University, Philadelphia, PA, USA. erjia.yan@drexel.edu
Lili, Lin, College of Computer and Information, Hohai University, China, linlili@hhu.edu.cn
Tamy Chambers, Department of Information and Library Science, Indiana University, Bloomington, IN, USA. tischt@imail.iu.edu

## Abstract
This paper proposes entitymetrics to measure the impact of knowledge units. Entitymetrics highlight the importance of entities embedded in scientific literature that for further knowledge discovery. In this paper, we use Metformin, a drug for diabetes, as an example to form an entity-entity citation network based on literature related to Metformin. We then calculate the network features and compare the centrality ranks of biological entities with results from Comparative Toxicogenomics Database (CTD). The comparison demonstrates the usefulness of entitymetrics to detect most of the outstanding interactions manually curated in CTD.

## Introduction
Currently, knowledge is being amassed rapidly; however, most of it is being encoded as strings in unstructured scientific literature. Extraction of this knowledge presently places a huge burden on already overloaded researchers, as they must manually dig out the embedded knowledge by reading tons of articles. This knowledge consists of many connected individual knowledge units encapsulated as entities in scientific papers. These entities could be authors, references, journals, and keywords that are commonly studied in scholarly evaluation, or they could be datasets, key methods, genes, drugs, and diseases that have not yet been widely explored in bibliometrics. Effective knowledge transfer depends on efficient knowledge accumulation. If these entities were decoded/annotated using a standard format (e.g. XML tags or RDF triples) and following shared semantics (e.g., domain ontologies, or controlled vocabularies), then connecting the entity dots would be as easy as flipping a switch. A knowledge graph could then be formed/accumulated automatically based on existing articles, and newly published articles, to lead potentially to successful knowledge discovery [1].

Articles have been an essential entity used in bibliometric studies for decades. This entity can be aggregated to measure journal, author, institution, and country/state level impact or divided to understand keyword use. Entities are either evaluative entities or knowledge entities. Evaluative entities are used to evaluate scholarly impact, including papers, authors, journals, institutions, and/or countries. Scholars have mainly studied these entities for two purposes: 1) evaluation of scholarly impact, such as identifying top influential players in a specific field using an author citation analysis (e.g., [2]), ranking prestigious journals using journal citation networks (e.g., [3]), or exploring social, cognitive, and geographic relationships between institutions through paper citation networks (e.g., [4]) and 2) examination of scientific collaboration behavior, such as analyzing the structure of scientific collaboration networks (e.g., [5]-[6]), mining patterns of author orders in scientific publications (e.g., [7]), or characterizing international scientific co-authorship patterns (e.g., [8]).

Knowledge entities act as carriers of knowledge units in scientific articles and include such entities as, keywords, topics, subject categories, datasets, key methods, key theories, and domain entities (e.g., biological entities: genes, drugs, and diseases). These knowledge entities are often to mine knowledge usage and transfer ultimately to facilitate knowledge discovery. Through co-word analysis, keywords have become a major knowledge unit used in current bibliometric analysis. However, they have limitations in detecting content interactions among scientific papers [9], portraying knowledge landscapes of specific domains [10] or science domains as a whole [11], and capturing existing schools of thoughts [12]. Recently, both subject categories and their upper level categories from the Web of Science (WOS) and/or Scopus have been used to analyze scientific trading between different domains [13].

The combination of evaluative entities and knowledge entities has been used to generate an overlay view of scholarly impact and knowledge usage to help interpret scholarly communication patterns through topical related explanation. Ding [14] combined evaluative entities (i.e., authors and papers) and knowledge entity (i.e. keywords) to explain whether productive authors tended to collaborate with and/or cite researchers with the same or different topical interests. Yan et al. [15] examined how research topics are mixed and matched in evolving research communities by using a hybrid approach to overlay keyword clusters and co-author networks. However, most bibliometric analyses use keywords as knowledge entities as they can be provided by WOS or Scopus and easily extracted from titles and abstracts.

Few bibliometric analyses have extended this knowledge entity to the domain entity level (e.g., genes, drugs, and diseases). In the biomedical domain, research often revolves around important bio-entities, such as diseases (e.g., Alzheimer's disease, Obesity, depression), drugs (e.g., metformin (Diabetes), troglitazone (Diabetes), Amitriptyline (Depression)), and genes (e.g., BRCA1 (Breast Cancer), APP (Alzheimer's disease), and LEP (Obesity)) [78]. Yet, current bibliometric analyses have not used these bio-entities (e.g., genes, drugs, and diseases) as knowledge entities. Analysis of citation relationships between bio-entities targeting a specific disease, drug, or gene could be used to provide in-depth understanding of knowledge usage and transfer in specific cases, and ultimately lead to knowledge discovery. Extracting knowledge these units is easier in the well-established and semantically stable domains, such as medicine, mathematics, geology, and finance. In these domains, controlled vocabularies and tools for extracting knowledge units have been developed by the community and are in common practice. Conversely, in the social sciences and humanities, where the semantics of knowledge units cannot be explicitly modeled and are highly contextualized, it can be challenging to apply entitymetrics.

As an example, Hammarfelt [79] used page citation analysis (PCA) to trace how different parts of the frequently cited publication - Walter Benjamin's Illuminations 91968/2007 - had been cited to study the intellectual structures of the humanities.

This paper proposes the new concept of entitymetrics (see Figure 1), which we define as using entities (i.e., evaluative entities or knowledge entities) in the measurement of impact, knowledge usage, and knowledge transfer to facilitate knowledge discovery. This extends bibliometrics by emphasizing the importance of entities, which are categorized as macro-level entities (e.g., author, journal, article), mesa-level entities (e.g., keyword), and micro-level entities (e.g., dataset, method, domain entities). These entities can be analyzed from the temporal perspective to capture dynamic changes or from the spatial dimension to identify geographical differences. Entitymetrics focused on both knowledge usage and discovery and can be viewed as the next generation of citation analysis [76], as it aims to demonstrate how bibliometric approaches can be applied to knowledge entities and ultimately contribute to knowledge discovery.

This paper uses Metformin, a drug for diabetes, as an example to illustrate the functionality and application of entitymetrics. Metformin was originally developed to treat Type II diabetes, but is now being considered in the treatment and prevention of cancer, obesity, depression, and aging [16]. Due to its significant drug repurposing function, Metformin has attracted great attention in diverse biomedical domains. This paper uses bio-entities as knowledge entities to analyze knowledge usage and transfer in Metformin related research. This paper is organized as follows: Section 2 outlines related work, Section 3 details the methods we applied in the present paper, Section 4 provides and discusses the research results, and Section 5 concludes the research and identifies future work.

## Related Work

*Bibliometric research using evaluative entities*
The common evaluative entities in bibliometrics are papers, authors, and journals. These entities can be aggregated to research groups, universities/institutions, countries, or disciplines. Van Raan [20] applied paper and author citation analysis to 147 university chemistry research groups in the Netherlands. He compared those with peer review judgments and found the results were correlated. Boyack et al. [11] mapped the structure of science and social science based on the journal citation networks and journal co-citation networks of 7,121 journals. H-index was proposed to combine the number of publications and number of citations to measure individual scientific achievement [21]. Co-author networks have been aggregated to the state and country levels to identify common patterns based on productivity and influence of author orders [7]. PageRank, and its variants, has been used to evaluate scientific impact, such as AuthorRank [22], Y-factor [3], CiteRank [23], FutureRank [24], Eigenfactor [25], and SCImago Journal Rank [26].

*Bibliometric research using knowledge entities*
Pettigrew and McKechnie [27] traced the usage of theories in information science (IS) based on 1,160 articles published in six information science journals. They analyzed both how authors applied IS theories in their published work and how those theories were used outside the field. They developed a code structure and manually coded each article. Through their analysis of these knowledge entities (e.g., IS theories), they were able to identify over 100 distinct theories born in IS and conclude that IS theory was not well cited outside the field. Yan et al. [13] used 221 Web of Science subject categories as knowledge entities to study knowledge transfer in the sciences and social sciences based on the journal citation networks. They found that the social science fields were becoming more visible by exporting more knowledge in scientific trading. Small [77] used the combined method of quantitative (clustering) and qualitative approaches (content analysis) to detect the interdisciplinary linkage between document clustering and journal subject categories.

Keyword is another important knowledge entity. Co-word analysis was first implemented in a system called LEXIMAPPE [28]. The co-word approach extracts keywords from articles and forms a co-occurrence matrix of these keywords. Callon, Courtial, and Laville [9] used keywords as knowledge entities to apply co-word analysis in the polymer science field. They identified the evolution of this field in different subject areas and demonstrated the research trajectories in a research network. Kostoff et al. [29] implemented another system called Database Tomography (DT), which extracted phrases and performed co-word analysis to detect relationships among themes and sub-themes. They demonstrated that co-word analysis could be used empirically to explain the emergence and diffusion of innovations [30]. However, the clustering method in the above co-word analyses is similar to the single linkage cluster algorithm, which is now considered unreliable [30]. Tijssen and Van Raan [31] showed that LEXIMAPPE could be replaced using multi-dimensional scaling (MDS). Leysdesdorff [32] pointed out several issues with LEXIMAPPE and proposed the use of factor analysis and other clustering techniques to conduct co-word analysis.

*Combination of evaluative entities and knowledge entities*
The combination of evaluative entities and knowledge entities can bring finer granularity to the ranking of evaluative entities by considering their contribution to concrete knowledge entities [33]. Ding [34] combined a knowledge entity (i.e.,topics) with an evaluative entity (i.e., author) to detect high impact authors in certain topics using topic-based PageRank. The Author-Conference-Topic (ACT) model, an extended Latent Dirichlet Allocation (LDA) model, has been used to extract topics and calculate topic distribution of individual authors and conferences [35]. Ding [14] analyzed scientific collaboration and citation networks by considering different topics as knowledge entities to determine whether productive authors tend to collaborate with, or cite, researchers with the same or different research topics and whether highly cited authors tend to collaborate with each other.

*Using entitymetrics to discover knowledge*
Swanson's work about undiscovered public knowledge has achieved a wide impact on association discovery and demonstrated that new knowledge can be discovered from sets of disjoint scientific articles [36]-[38]. Swanson [39] pointed out that bibliometrics can be a valuable tool for knowledge discovery given that it analyzes the citing and cited relationships of articles. Swanson further suggested that his method could be extended to many other disconnected literature sets to enable cross-disciplinary innovation [40]. Arrowsmith (http://arrowsmith.psych.uic.edu/arrowsmith_uic/index.html) was thus developed to enable semi-automatic knowledge discovery [41]. The basic assumption of Swanson's method is that knowledge developed in one field maybe unknown in another field. There are three ways to ensure the mutual exclusion for two literature sets [41]-[42]: 1) by excluding common papers which appear in both sets (they called A set and C set), 2) and even stricter, by excluding papers from both sets which cite the same papers, and 3) the strictest, by making sure that papers from both sets have not been co-cited together before. There are several problems with Swanson's method. First, it is not easy to identify A and C directly without domain specific knowledge. For example, the popular example for Swanson's method uses Raynaud's Disease (A) and eicosapentaenoic acid (C), or magnesium deficiency (A) and migraine (C), which took years to identify hypotheses to test the connections between these specific diseases and drugs. Second, it is difficult to identify important terms connecting A and C sets (called B-terms) as there could be thousands of B-terms that link both sets. The rank of B-terms based on frequency, calculated from a Poisson distribution, is far from sufficient to identify potential breakthroughs. Usually the top-ranked B-terms, based on frequency, are known knowledge, while the low-ranked B-terms are overwhelming in number and noisy. Gordon and colleagues improved the Swanson system by applying a statistical method to rank B-terms [43]-[44]. Weeber et al. [45] improved Swanson's model by converting terms to knowledge entities using the biomedical Unified Medical Language System (UMLS) concepts as units of analysis. Weeber's system has successfully simulated Swanson's discoveries of connecting Raynaud's disease with fish oil and migraine with magnesium deficiency thereby demonstrating the main advantage of using UMLS concepts over keywords. In doing so, they extended literature-based discovery to data-based discovery by involving databases, such as annotated genetic databases. Stegmann and Grohmann [46] applied Swanson's method to generate hypothesis for Raynaud's Disease that demonstrated co-word clustering as a powerful method for literature-based hypothesis generation and knowledge discovery. Bekhuis [47] summarized all developments and applications based on the early work of Swanson and claimed that Swanson's vision of the hidden value in the literature of science in biomedical digital databases is remarkably innovative for information scientists, biologists, and physicians.

## Methodology

*Metformin*

Metformin ($N''$,$N''$-dimethylbiguanide) is often referred to as a magic drug. Originally, the drug was developed to treat Type II diabetes, however now, it is also being considered to treat and prevent cancer, obesity, depression, and aging [16]. Endocrinologists, cardiologists, oncologists, and aging specialists have generated waves of interest by attempting to use anti-diabetic biguanides to control body weight and tumor growth [48]. Metformin is the only anti-diabetic drug, which can prevent the cardiovascular complications of diabetes and remains as one, of the only two, oral anti-diabetics on the World Health Organization Model List of Essential Medicines [49]. Obesity and cancer have interrelationships with aging [50]. Metformin is able moderately to reduce body weight for obese diabetics because it can reduce insulin resistance and hyperinsulinemia [51]. A recent study found that children and adolescents are more responsive to metformin-induced weight loss compared to adults [52].

*Dataset*

PubMed and PubMed Central (PMC) are used to generate a biological entity citation network for Metformin related articles. PubMed Central is the U.S. National Institutes of Health (NIH) digital archive of full-text biomedical and life sciences journal literature. Our collection from PMC contains 353,592 articles from 3,068 journals and our dataset from PubMed contains 20,494,848 papers published between the years 1966 and 2011. The dataset was used to extract citation relationships between papers with the criteria that both citing papers and cited papers have PubMed IDs (PMID) so that bio-entities can be connected via citation.

*Biological Entity Extraction*
To identify entities in all PubMed articles, we employed a dictionary-based named entity recognition method with exact match. The dictionary, taken from Wang et al. [53], includes three parts: genes, diseases, and drugs. The dictionary is built from a Drug dictionary (DrugBank (http://www.drugbank.ca/)), a Target dictionary (HUGO (http://www.genenames.org/)), and a Disease dictionary (MESH disease from CTD (http://ctdbase.org/)). A small chunk of the dictionary is shown in Figure 2. In the example entry (ACTA1-->GENE__P68133", "actin, alpha 1, skeletal muscle-->GENE__P68133), ACTA1 is the name of the gene and P68133 is its MeshID. The example shows that, besides searching for "ACTA1" in the document, all the other synonyms or aliases (actin, alpha 1, and skeletal muscle) are also searched. The extraction was conducted on the title and abstract of each publication using the LingPipe library from the Alias–i project, which contains a package called Exact Dictionary-Based Chunking. After finishing the extraction, all the results were stored in a relational database to facilitate further processing.

*Entity Citation Network*
In the next step, we built a bio-entity citation network based on the concept that if paper A cites paper B, then an entity in paper A will be considered to cite an entity in paper B. Entities in the citing paper are paired with entities in all the cited papers (see Figure 3). A hash table is used to store the entity citation associations and their occurance frequency.

Figure 4 illustrates the process used to create the entity citation network which includes three components: ArticleFilter, EntityFetcher, and GraphCreator. The ArticleFilter component extracts a set of references from the reference section of papers related to a target object (e.g., a diseasese, a concept, and a method), which are showed in squared parenthesis. Subsequently, the EntityFetcher component retrieves entities for this set of references. Finally, the GraphCreator component generates a hash table of entity citation relationships and counts the number of times each relationship occurs. In the final graph, vertices represent entities and edges represent citation relationships with number of citations as weights. In this paper, the ArticleFilter is applied to get the list of references from the PubMed papers related to Metformin, then the EntityFetcher collected extracted entities from this list of references, finally the GraphCreator generated a entity citation graph based on the entities retrieved from the EntityFetcher and citation relationships captured by the ArticleFilter.

*Metformin-related Entity Citation Network*
In order to guarantee the coverage of Metformin related articles, search terms were extended from only Metformin to include brand name and synonyms, as well as, related diseases and genes extracted from the CTD and DrugBank (see Table 1).

The following search terms were used to search the downloaded PubMed Central dataset from NIH: ("metformin"[ti] OR "metformin"[ab] OR "alstrom syndrome"[ti] OR "alstrom syndrome"[ab] OR "amyloidosis"[ti] OR "amyloidosis"[ab] OR "atrophy"[ti] OR "atrophy"[ab] OR "diabetes complications"[ti] OR "diabetes complications"[ab] OR "diabetes mellitus"[ti] OR "diabetes mellitus"[ab] OR "diabetes mellitus (experimental)"[ab] OR "diabetes mellitus (type 2)"[ti] OR "diabetes mellitus (type 2)"[ab] OR "diabetic angiopathies"[ab] OR "diabetic nephropathies"[ab] OR "hyperandrogenism"[ti] OR "hyperandrogenism"[ab] OR "hyperglycemia"[ti] OR "hyperglycemia"[ab] OR "hyperinsulinism"[ti] OR "hyperinsulinism"[ab] OR "hypertension"[ti] OR "hypertension"[ab] OR "insulin resistance"[ti] OR "insulin resistance"[ab] OR "myocardial infarction"[ti] OR "myocardial infarction"[ab] OR "obesity"[ti] OR "obesity"[ab] OR "polycystic ovary syndrome"[ti] OR "polycystic ovary syndrome"[ab] OR "albuminuria"[ti] OR "albuminuria"[ab]) AND ("1965/01/01"[PubDate] : "2011/12/31"[PubDate]). Finally, of the 4,770 articles retrieved from PubMed Central (the citing article set), only those references with PubMed IDs were kept (the cited article set), which resulted in 134,844 references. The references without PubMed IDs were not included, as bio-entities could not be extracted from these references. From the titles and abstracts of 4,770 full text articles, 1,969 bio-entities (i.e., 880 genes, 376 drugs, and 713 diseases) were extracted. Table 2 shows the top 20 ranked bio-entities.

In the cited article set (e.g., 134,844 references), 6,978 entities were extracted including 3,914 genes, 1,296 drugs, and 1,768 diseases. Table 3 shows the top 20 ranked bio-entities, which are highly cited.

*Network Analysis Approaches*
Network analysis approaches consist of three levels: macro-level statistics (global graph metrics), meso-level structures (cluster characters), and micro-level indicators (individual node properties). The macro-level analyses includes component, bi-component, *k*-core, shortest distance, and degree distribution; the meso-level analyses mainly includes clustering coefficient, as in the current study, but may also include techniques such as, hierarchical clustering and modularity-based clustering; and the micro-level analyses refer to different centrality measures. For formal definitions of these approaches, readers can refer to Freeman [54]-[56] and Nooy, Mrvar, and Batagelj [57].

*Macro-level features*
**Component analysis:** In network analysis, connected graphs are called components. Component analysis can be used to learn about the macro-level structure of a network. Some network analysis methods (e.g., shortest distance) can only be used in connected networks, and thus, are only applied to the largest component.

**Bi-component analysis:** In a bi-component graph, no node can control the information flow between two other nodes completely because there is always an alternative path that information may follow [57]. In a bi-component graph, each node receives information from at least two nodes. In these bi-components graphs, nodes share similar information and are identical to each other [58].

**K-core analysis:** The k-core of a network is a sub-structure in which each node has ties to at least k other nodes [59]. Nodes in the core are tightly linked, thus ties in each k-core are strong ties [60]. Information transfer within a k-core maybe redundant, as one vertex can receive the same information more than once from other nodes in the same k-core.

**Distance analysis:** A geodesic is the shortest path between two nodes. Many networks show that most individuals are at very few "degrees of distance" from one another [61]. The mean shortest distance between node pairs in a network can be expressed as:

$$L = \frac{1}{\frac{1}{2}N(N+1)} \sum_{i \geq j} d_{ij}$$

where $d_{ij}$ is the geodesic distance from node i to node j; and N is the total number of nodes in the connected component.

**Degree distribution:** The degree of a node is the number of other nodes connected to it. Nodes with higher degrees act as hubs in the network and are crucial to the robustness of the network, as well as, the flow of information. Degree distribution measures the character of a network: a few nodes have many links and a majority have a smaller numbers of links. Albert and Barabási [62] discovered that degree distribution in many real-world networks follows the power-law distribution: $p(k) \sim k^{-\gamma}$ where k is the node degree and $\gamma$ is a constant.

*Meso-level features*

Networks contain local communities/clusters in which many nodes are "locally" connected with one another [5] [63]. Clustering coefficient is an effective meso-level indicator to estimate the locally clustering feature:

$$C = \frac{3 \times number\,of\,triangles\,on\,the\,graph}{number\,of\,connected\,triples\,of\,vertices}$$

This definition corresponds to the concept of the "fraction of transitive triples" used in sociology [64].

*Micro-level features*

**Degree centrality:** Degree centrality for a node is the number of links that a node has with others, which can be expressed as follows:

$$C_D(n_i) = d(n_i)$$

where $d(n_i)$ is the degree of $n_i$.

**Closeness centrality:** Unlike degree centrality, the closeness centrality of a node focuses on its extensibility of influence over the entire network and is expressed as:

$$C_c(n_i) = \sum_{i=1}^{N} \frac{1}{d(n_i, n_j)}$$

where $C_c(n_i)$ is the closeness centrality, and $d(n_i, n_j)$ is the distance between two nodes in the network.

**Betweenness centrality:** Betweenness centrality is based on the number of shortest paths passing through a node. Nodes with a high betweenness serve as bridges that connect different sub-groups. Betweenness is expressed as:

$$C_B(n_i) = \frac{\sum_{j<k}(n_i)}{g_{jk}}$$

where $g_{jk}$ is the geodesic distance between the nodes of j and k.

*Evaluation*

For evaluation, we use the CTD (Comparative Toxicogenomics Database) which contains 384,141 chemical synonyms, 679,701 gene synonyms, and 68,211 disease synonyms. The CTD provides us with a set of 336,693 interactions between diseases and drugs. To measure the significance of Metformin in disease, we utilize the inference score, provided in the CTD, to measure the strength of the association between diseases and Metformin or its descendants. Each association is curated as either marker/mechanism or therapeutic, which is used for calculating the inference score. The inference score, proposed by King et al. [66], is used to represent the similarity of chemical–gene–disease networks in the CTD by comparing a scale-free random network with a similar topology. The higher the inference score, the more likely the inference network will show distinct linkage.

## Result and Discussion

In total, the entity-entity citation network based on Metformin information contains 7,055 entities and 264,345 links, among which 1,561 are loops (i.e. self-citations).

*Macro-level features*

All of the 7,055 entities belong to exactly one component and bi-component. This means that the network is well connected, as at least two distinct semi-paths connect every pair of entities. The density of a network shows the degree of connections between any given two pairs in this network. If it is applied to data with values, density shows the average strength of the ties across all possible ties. The density of the Metformin network is 0.005311, which means that 0.5311% of all possible connections are presented in the current network. According to the K-core analysis, the biggest k-core (188-core) consist of 238 entities, which means that each entity in this sub-network has a connection with at least 188 others. The mean geodesic distance is 2.10. This means the average of shortest path between any two nodes is about two nodes long (not including the two given nodes). Therefore, information can be transferred efficiently through this network. The diameter (e.g., the largest geodesic distance between nodes pairs in the network) is four; between GENE otc and GENE ube2v1. This indicates that there is a close relation among all the entities, as every pair of entities could be reached by one another within three steps. Figure 5 shows the longest path (e.g., the diameter of the network) from Gene otc to Gene ube2v1.

As the entity-entity citation network is a direct network, histograms can be used to show in-degree and out-degree distributions (Figure 6). A power law distribution was found in both the in-degree and out-degree distributions. This means that a low portion of the nodes have a high in-degree/out degree links, while a majority of the nodes have very few in-degree/out-degree links. The results confirm Albert's [65] study that most cellular interaction networks are scale-free.

*Meso-level features*

The clustering coefficient of the network was 0.684687, which indicates that entities in the Metformin network have a high tendency to cluster together. In Newman's [63] review article, the author highlighted the properties of a few biological networks, including metabolic networks, protein interactions, marine food web, freshwater food web, and neutral network. The networks all have smaller average shortest paths (ranging from 1.9 to 6.8) and clustering coefficients (ranging from 0.09 to 0.20). The current network also exhibits these small-world properties including, a small average shortest path (2.1) and a large clustering coefficient (0.68).

*Micro-level features*

The degree (in-degree and out-degree) centrality, closeness centrality, and betweenness centrality of this network were calculated. Entities with high centrality are listed in Tables 4-7 which contain the top 20 ranks for each of three different kinds of entities (e.g., disease, drug, and gene) and for all the entities combined.

There are 697 diseases ranked for Metformin in the CTD. In terms of in-degree centrality, we found 16 matches including the four common terms: disease, syndrome, death, and body weight. The following three diseases ranked in top 10: Diabetes mellitus (1st), Obesity (7th), and Insulin resistance (9th). The six diseases ranked between 11th and 100th: Atherosclerosis (20th), Hypertension (21st), Myocardial infarction (28th), Inflammation (37th), Heart Failure (40th), and Stroke (74th). The following seven diseases ranked low: Necrosis (102nd), Ischemia (120th), Coronary artery disease (174th), Infection (265th), Erythema (306th), Hyperglycemia (337th), and Hypertrophy (439th).

Since the CTD curates specific drug-gene interactions in vertebrates and invertebrates from the published literature, we utilize this information to understand the importance of the genes identified in terms of in-degree centrality. Among the 20 genes we identified, the following five matched with the CTD entries: TNF, LEP, JUN, CAT, and Glucagon. The most salient gene that interacts with Metformin is TNF. We found ten interactions between Metformin and TNF (ranked 3rd in the CTD). Studies have shown Metformin increases expression of the TNS protein, leading to improved hepatic steatosis, when co-treated with rosiglitazone [67] and that Metformin reduces Streptozocin which thus results in an increased expression of TNF mRNA[68]-[69]. Regarding the interaction between Metformin and LEP, we found only one interaction consistent with reports that Metformin results in an increased expression of the LEP gene [70]. Between Metformin and JUN, we found only one interaction that Metformin inhibits Tetradecanoylphorbol Acetate, which affects the localization of JUN gene [71]. Between Metformin and CAT gene, we found three interactions stating that Metformin inhibits Streptozocin thus resulting in decreased expression of CAT mRNA [68]. Between Metformin and Glucagon, there are four interactions indicating that Metformin results in increased expression of IL1RN mRNS when co-treated with Glucagon [72].

Since the CTD does not provide information about the interaction between Metformin and other drugs, we consulted with a well-known drug interaction checker (http://www.drugs.com/drug-interactions). A drug interaction means that another drug affects the activity of the drug when both are administered together. This interaction can be either synergistic or antagonistic; and could sometimes produce effects not achieved by either drug individually. On the drug interaction checker website, we found 14 drugs to have a major interaction with Metformin and 589 drugs to have a moderate interaction with Metformin. Among the top 20 drugs identified in terms of in-degree centrality, the following four were identified as having a moderate interaction with Metformin according to the drug interaction checker: Dopamine, Dexamethasone, Ethanol, and Epinephrine. Dopamine is a monoamine neurotransmitter and hormone that plays a significant role in the body of animals. Dexamethasone is a steroid drug that acts as an anti-inflammatory. Ethanol is a psychoactive drug that leads to a state of alcohol intoxication when consumed. Epinephrine is a hormone that carries out many crucial functions in the body such as regulating heart rate, blood vessel, etc. Both Dopamine and Epinephrine have a synergistic relationship with Metformin by reducing its effects in lowering the blood sugar, whereas Dexamethasone has an antagonistic relationship with Metformin by causing a condition called lactic acidosis.

Using out-degree centrality, we found 16 matched entries for disease. Compared to in-degree centrality, 14 diseases are identical between in-degree and out-degree. Ischemia and coronary artery disease are not ranked in the top 20 for out-degree centrality. Three disease interactions, fibrosis, atrophy, and cardiovascular diseases, are newly ranked in top 20 for out-degree centrality. In the CTD, fibrosis ranked 54th, cardiovascular disease ranked 112th, and Atrophy ranked 215th.

For the interaction between Metformin and genes, out-degree centrality identifies five matched genes with the CTD: LEP, TNF, MMP9, JUN, and CRP. LEP and TNF are both explained above. We found four interactions between Metformin and MMP9 in the CTD. A study has shown that Metformin inhibits Tetradecanoylphorbol Acetate and results in increased activity of MMP9 gene [71]. We found that there are three interactions between Metformin and CRP consistent with reports that Metformin results in decreased expression of the CRP gene [73]-[74].

In the interaction between Metformin and other drugs, we found two matches with the drug interaction checker, Dopamine and Ethanol. Both are identified by in-degree centrality and are explained earlier.

Using closeness centrality, we found 13 matches with no new interactions between Metformin and diseases identified. Among the 13 matches, the following three are in top 10: Diabetes mellitus (1st), Obesity (7th), and Insulin resistance (9th). Five diseases ranked between 11th and 100th including: Hypertension (21st), Myocardial infarction (28th), Inflammation (37th), Heart Failure (40th), and Stroke (74th). The following six diseases ranked low: Necrosis (102nd), Atrophy (215th), Infection (265th), Erythema (306th), Hyperglycemia (337th), and Hypertrophy (439th). With respect to the interaction between Metformin and other drugs and the interaction between Metformin and genes, the closeness centrality identified four interactions: TNF, LEP, Glucagon, and JUN. There are no new interactions in top 20 genes found by closeness centrality.

Using the betweenness centrality, we found 17 Metformin and disease interactions. The following three diseases ranked in the top 10: Diabetes mellitus (1st), Obesity (7th), and Insulin resistance (9th). Seven diseases ranked between 11th and 100th including: Atherosclerosis (20th), Hypertension (21st), Myocardial infarction (28th), Inflammation (37th), Heart Failure (40th), Fibrosis (54th), and Stroke (74th). The following seven diseases ranked low: Necrosis (102nd), Ischemia (120th), Atrophy (215th), Infection (265th), Erythema (306th), Hyperglycemia (337th), and Hypertrophy (439th).

The following four genes interaction with Metformin are identified by betweenness centrality: TNF, LEP, MMP9, and Glucagon (all genes are explained earlier). For Metformin and drug interaction, the following four drugs are identified: Dopamine, Dexamethasone, Testosterone, and Epinephrine. The newly identified drug by betweenness centrality, Testosterone, is a steroid hormone that plays an imperative role in developing male reproductive tissues such as the testis and prostate. Like the other three drugs, testosterone reduces the effects of Metformin by lowering the blood sugar.

# Conclusion

This paper proposes entitymetrics to measure the impact of knowledge units at different levels. It highlights the importance of entities embedded in scientific literature for further knowledge discovery. Compared with the related work, entitymetrics advances the state of the art by taking knowledge entity as the research unit to move bibliometrics to discovery knowledge. It refines Swanson's method by utilizing B terms as knowledge entities. This paper uses Metformin as an example to form an entity-entity citation network based on literatures related to Metformin and calculates the network features of this network. It compares the centrality ranks of the network with results from the CTD. Entitymetric results, identifying the interaction of Metformin with diseases, drugs, and genes, were consistent with the CTD thereby demonstrating the usefulness of the entity level bibliometric approach to detect most of the outstanding interactions manually curated in the CTD.

The results also show that our approach is complimentary to CTD. The CTD reported that 124 genes interact with metformin. Out of the genes identified by our meso-level analysis, we identified 30 unique gene interactions with metformin, and among these 30 genes, eight matched with the CTD. We conducted a literature review to find whether the 22 unmatched genes interact with metformin. Table 8 shows the list of the genes found to be in interaction with metformin.

Because of manual curation, the CTD can provide only a limited coverage of interactions among bio-entities. Therefore, interactions identified by our approach, but which are not found in the CTD, are not necessarily insignificant, but rather may indicate a novel interaction not previously reported and worthy of further investigation. For example, a recent study by Elia et al. [80] shows that Metformin changes the peroxisome proliferator-activated receptor in the uterine tissue of mice. This interaction between Metformin and peroxisome proliferator-activated receptor is not reported in the CTD, but our approach identified an interaction between these two entities. Another example is our identification of an interaction between metformin and resistin which is supported by the recent discovery of Labuzek et al. [70] that Metformin treatment had a positive impact on up-regulating resistin. Explained by these examples, we expect our approach could infer other potential interactions which could later be confirmed by clinical experiments.

Literature-based knowledge discovery aims to connect the disconnected scientific entities to generate new knowledge. Although data-based knowledge discovery is based on more stringently validated data from experiments or clinical trials, the benefit of literature-based discovery can also be enormous. The connections between concepts in scientific literature can co-occur if two concepts are co-occur in a predefined context (e.g., title, abstract, one sentence, or one paragraph), can cite if the paper mentioning concept A cites the paper discovering/discussing concept B, and can co-cite if the paper mentioning concept A and the paper containing concept B are co-cited by other articles.

Divided specialization fragments science and disconnects adjacent disciplines. Scientific collaboration glues science back together and connects these disconnections. Scientific articles are co-authored based on needed expertise and interlinked through citations. Today, data exists in diverse formats (e.g., textual, visual, and numeric) and are available in technical reports, clinical trials, gene or protein sequence databases, patient records, medical device recordings, and sensor recordings. Dotted knowledge can be connected by mining data across boundaries. For example, co-author connections in articles can reflect scientific collaboration patterns and gene co-occurrence connections in articles can identify potential association between genes. The overlay of co-author networks with gene co-occurrence networks can portray the entity-oriented scientific collaboration landscapes.

The problem solving style in the biomedical domain is diagnostic. Generating and testing hypotheses are traditions in scientific discovery in the biomedical domain. Given the fast growth in scientific literature, literature-based approaches for generating hypotheses are quickly emerging. Blagosklonny and Pardee [75] proposed that conceptual biology should take advantage of millions of accumulated data in databases and a variety of sources from thousands of journal articles to generate new knowledge "by reviewing these accumulated results in a concept-driven manner, linking them into testable chains and networks" (p.373). Mining and connecting biological entities in published articles can integrate unknown knowledge and should work closely with lab experimental verification. As Swanson [39] argued "neither relationship by itself is necessarily of much interest, but two literatures that are both non-interactive and yet logically related may have the extraordinary property of harboring undiscovered causal connections" (p. 131). As mentioned in the Related Work section, it is hard to apply Swanson's method without knowing the A and C terms. The recent test conducted by us shows some promising results that co-word analysis can be used to identify potential A or C terms. Citation analysis can unveil the disconnected knowledge and co-citation can discover implicit knowledge connections. The combination of both could help us to develop a synthetic mechanism to enable knowledge discovery, which could strike up new developments and applications.

## Acknowledgement

This study was supported by National Research Foundation of Korea Grant funded by the Korean Government (NRF-2012-2012S1A3A2033291) and by the Bio & Medical Technology Development Program of the National Research Foundation (NRF) funded by the Korean government (MEST) (2012048758).

**Figure Legends**

Figure 1: Entitymetrics
Figure 2. A fraction of bio-entity dictionary
Figure 3. From paper citation to entity citation
Figure 4. A overview of Entity Citation Network generation
Figure 5. The longest path of the network

Figure 6. In-degree and out-degree distribution of the Metformin entity-entity citation network

**Tables**

**Table 1. Metformin naming variations and related Genes and Diseases**

| Brand Name | Fortamet, Apo-Metformin, Gen-Metformin, Glucophage, Glucophage XR, Glumetza, Glycon, Mylan-Metformin, Novo-Metformin, Nu-Metformin, PMS-Metformin, Ran-Metformin, Ratio-Metformin, Riomet, Sandoz Metformin, Teva-Metformin |
|---|---|
| Synoym | LA-6023 |
| ACT Code | A10BA02 |
| IUPAC name | 1-carbamimidamido-N,N-dimethylmethanimidamide |
| Target (Three Synonyms) | 5'-AMP-activated protein kinase subunit beta-1, AMPK beta-1 chain, AMPKb |
| CasRN | 657-24-9 |
| Related Diseases | Alstrom Syndrome, Amyloidosis, Atrophy, Diabetes Complications, Diabetes Mellitus, Diabetes Mellitus (Experimental), Diabetes Mellitus(Type 2), Diabetic Angiopathies, Diabetic Nephropathies, Hyperandrogenism, Hyperglycemia, Hyperinsulinism, Hypertension, Insulin Resistance, Myocardial Infarction, Obesity, Polycystic Ovary Syndrome, Albuminuria |
| Related Genes | CASP3, CASP7, CASP8, FASN, HRH2, MMP2, MMP9, NR1I3, PLAT, PRKAA1, PRKAA2, SLC15A1 |

**Table 2. Top 20 ranked bio-entities with high frequency in the citing article set**

| Frequency | Entity | Entity ID | Entity Name |
|---|---|---|---|
| 1458 | DISEASE | D009765 | Obesity |
| 1213 | DISEASE | D006973 | Hypertension |
| 989 | DISEASE | D004194 | Disease |
| 670 | DISEASE | D003920 | diabetes mellitus |
| 622 | DISEASE | D007333 | insulin resistance |
| 555 | DISEASE | D009203 | myocardial infarction |
| 552 | GENE | P01308 | Insulin |
| 362 | DISEASE | D001284 | Atrophy |
| 351 | DISEASE | D013577 | Syndrome |
| 275 | DISEASE | D006943 | Hyperglycemia |
| 257 | DISEASE | D007249 | Inflammation |
| 249 | GENE | Q9UH22 | Large |
| 205 | GENE | Q9P2X3 | Impact |
| 193 | DISEASE | D050177 | Overweight |
| 189 | DISEASE | D003643 | Death |
| 162 | DISEASE | DB04540 | Cholesterol |
| 141 | DISEASE | D050197 | Atherosclerosis |
| 137 | DISEASE | D001835 | body weight |

| 137 | DISEASE | D020521 | Stroke |

**Table 3. Top 20 highly cited bio-entities**

| Num. of Papers | Entity | Entity ID | Entity Name |
|---|---|---|---|
| 22867 | DRUG | DB04077 | Glycerol |
| 22846 | DRUG | DB04557 | arachidonic acid |
| 22511 | DISEASE | D004890 | Erythema |
| 19978 | DISEASE | D004194 | Disease |
| 13323 | GENE | P01308 | Insulin |
| 11332 | DISEASE | D009765 | Obesity |
| 10422 | DISEASE | D006973 | Hypertension |
| 6583 | DISEASE | D007333 | insulin resistance |
| 5769 | DISEASE | D003920 | diabetes mellitus |
| 5630 | DISEASE | D013577 | Syndrome |
| 5604 | GENE | Q9UH22 | Large |
| 4721 | DISEASE | D003643 | Death |
| 4347 | DISEASE | D007249 | Inflammation |
| 4289 | DRUG | DB04540 | Cholesterol |
| 4023 | DISEASE | D009203 | myocardial infarction |
| 3294 | DISEASE | D001835 | body weight |
| 3278 | DISEASE | D050197 | Atherosclerosis |
| 3211 | GENE | Q9P2X3 | Impact |
| 2864 | DRUG | DB00435 | nitric oxide |
| 2427 | DRUG | DB01373 | Calcium |

**Table 4. In-degree centrality (top 20)**

| Rank | Disease | Drug | Gene | All Entities |
|---|---|---|---|---|
| 1 | DISEASE_disease | DRUG_glycerol | GENE_large | DISEASE_disease |
| 2 | DISEASE_erythema | DRUG_arachidonic acid | GENE_insulin | DRUG_glycerol |
| 3 | DISEASE_syndrome | DRUG_calcium | GENE_impact | DISEASE_erythema |
| 4 | DISEASE_death | DRUG_cholesterol | GENE_set | DRUG_arachidonic acid |
| 5 | **DISEASE_hypertension** | DRUG_nitric oxide | **GENE_tnf** | GENE_large |
| 6 | **DISEASE_obesity** | DRUG_potassium | **GENE_lep** | DISEASE_syndrome |
| 7 | **DISEASE_inflammation** | DRUG_glutathione | GENE_hr | DISEASE_death |
| 8 | **DISEASE_diabetes mellitus** | DRUG_ester | GENE_ca2 | GENE_insulin |
| 9 | **DISEASE_necrosis** | **DRUG_dexamethasone** | GENE_camp | GENE_impact |
| 10 | **DISEASE_insulin resistance** | DRUG_norepinephrine | GENE_met | **DISEASE_hypertension** |
| 11 | DISEASE_body weight | DRUG_zinc | GENE_rest | **DISEASE_obesity** |
| 12 | **DISEASE_atherosclerosis** | DRUG_heparin | GENE_albumin | **DISEASE_inflammation** |
| 13 | **DISEASE_infection** | DRUG_acetylcholine | GENE_renin | **DISEASE_diabetes mellitus** |
| 14 | **DISEASE_stroke** | **DRUG_Dopamine** | GENE_insulin receptor | GENE_set |
| 15 | **DISEASE_hypertrophy** | DRUG_iron | **GENE_glucagon** | DISEASE_necrosis |

| 16 | **DISEASE_myocardial infarction** | DRUG_aldosterone | GENE_plasminogen | **DISEASE_insulin resistance** |
| 17 | **DISEASE_heart failure** | DRUG_adenosine | **GENE_jun** | DRUG_calcium |
| 18 | **DISEASE_hyperglycemia** | DRUG_l-arginine | GENE_myoglobin | DISEASE_body weight |
| 19 | **DISEASE_ischemia** | **DRUG_epinephrine** | **GENE_cat** | **DISEASE_atherosclerosis** |
| 20 | **DISEASE_coronary artery disease** | DRUG_creatine | GENE_tg | DRUG_cholesterol |

**Table 5. Out-degree centrality (top 20)**

| Rank | Disease | Drug | Gene | All Entities |
|---|---|---|---|---|
| 1 | DISEASE_disease | DRUG_cholesterol | GENE_insulin | DISEASE_disease |
| 2 | **DISEASE_obesity** | DRUG_calcium | GENE_large | **DISEASE_obesity** |
| 3 | **DISEASE_hypertension** | DRUG_nitric oxide | GENE_impact | **DISEASE_hypertension** |
| 4 | **DISEASE_diabetes mellitus** | DRUG_aldosterone | **GENE_lep** | **DISEASE_diabetes mellitus** |
| 5 | **DISEASE_insulin resistance** | DRUG_glycerol | **GENE_tnf** | **DISEASE_insulin resistance** |
| 6 | DISEASE_atrophy | DRUG_arachidonic acid | GENE_renin | GENE_insulin |
| 7 | DISEASE_syndrome | DRUG_metformin | GENE_insulin receptor | **DISEASE_atrophy** |
| 8 | **DISEASE_inflammation** | DRUG_potassium | GENE_set | DISEASE_syndrome |
| 9 | **DISEASE_myocardial infarction** | DRUG_zinc | **GENE_mmp9** | **DISEASE_inflammation** |
| 10 | **DISEASE_hyperglycemia** | DRUG_fructose | **GENE_mmp2** | GENE_large |
| 11 | DISEASE_death | DRUG_norepinephrine | GENE_resistin | **DISEASE_myocardial infarction** |
| 12 | **DISEASE_atherosclerosis** | DRUG_rosiglitazone | **GENE_glucagon** | **DISEASE_hyperglycemia** |
| 13 | **DISEASE_stroke** | DRUG_glutathione | GENE_ace | DISEASE_death |
| 14 | **DISEASE_infection** | **DRUG_Dopamine** | GENE_pah | GENE_impact |
| 15 | DISEASE_body weight | DRUG_pioglitazone | GENE_peroxisome proliferator-activated receptor gamma | **DISEASE_atherosclerosis** |
| 16 | **DISEASE_heart failure** | DRUG_sildenafil | GENE_cd4 | **DISEASE_stroke** |
| 17 | **DISEASE_fibrosis** | **DRUG_ethanol** | **GENE_crp** | **DRUG_cholesterol** |
| 18 | **DISEASE_hypertrophy** | DRUG_iron | GENE_albumin | **DISEASE_infection** |
| 19 | **DISEASE_cardiovascular diseases** | DRUG_urea | GENE_hr | DISEASE_body weight |
| 20 | **DISEASE_necrosis** | DRUG_sucrose | GENE_rhodopsin | **DISEASE_heart failure** |

**Table 6. Closeness centrality (top 20)**

| Rank | Disease | Drug | Gene | All Entities |
|---|---|---|---|---|
| 1 | DISEASE_disease | DRUG_glycerol | GENE_insulin | DISEASE_disease |
| 2 | **DISEASE_obesity** | DRUG_arachidonic acid | GENE_large | **DISEASE_obesity** |
| 3 | **DISEASE_hypertension** | DRUG_cholesterol | GENE_impact | **DISEASE_hypertension** |
| 4 | **DISEASE_diabetes mellitus** | DRUG_calcium | GENE_set | GENE_insulin |
| 5 | DISEASE_syndrome | DRUG_nitric oxide | **GENE_tnf** | **DISEASE_diabetes mellitus** |
| 6 | **DISEASE_insulin resistance** | DRUG_aldosterone | **GENE_lep** | DISEASE_syndrome |
| 7 | **DISEASE_atrophy** | DRUG_potassium | GENE_renin | **DISEASE_insulin resistance** |
| 8 | **DISEASE_inflammation** | DRUG_zinc | GENE_insulin receptor | GENE_large |

| 9 | DISEASE_death | DRUG_glutathione | GENE_hr | **DISEASE_atrophy** |
|---|---|---|---|---|
| 10 | **DISEASE_myocardial infarction** | DRUG_norepinephrine | GENE_camp | **DISEASE_inflammation** |
| 11 | **DISEASE_hyperglycemia** | **DRUG_Dopamine** | **GENE_glucagon** | DISEASE_death |
| 12 | **DISEASE_erythema** | DRUG_iron | GENE_myoglobin | **DISEASE_myocardial infarction** |
| 13 | **DISEASE_atherosclerosis** | DRUG_rosiglitazone | GENE_albumin | GENE_impact |
| 14 | **DISEASE_infection** | DRUG_ester | GENE_plasminogen | **DISEASE_hyperglycemia** |
| 15 | **DISEASE_stroke** | DRUG_fructose | GENE_cd4 | DRUG_glycerol |
| 16 | DISEASE_body weight | DRUG_metformin | **GENE_jun** | DRUG_arachidonic acid |
| 17 | **DISEASE_necrosis** | **DRUG_dexamethasone** | GENE_peroxisome proliferator-activated receptor gamma | **DISEASE_erythema** |
| 18 | **DISEASE_heart failure** | DRUG_l-arginine | GENE_rest | **DISEASE_atherosclerosis** |
| 19 | **DISEASE_hypertrophy** | DRUG_urea | GENE_ace | **DISEASE_infection** |
| 20 | **DISEASE_fibrosis** | DRUG_adenosine | GENE_resistin | **DISEASE_stroke** |

**Table 7. Betweenness centrality (top 20)**

| Rank | Disease | Drug | Gene | All Entities |
|---|---|---|---|---|
| 1 | DISEASE_disease | DRUG_glycerol | GENE_large | DISEASE_disease |
| 2 | **DISEASE_obesity** | DRUG_arachidonic acid | GENE_insulin | **DISEASE_obesity** |
| 3 | **DISEASE_hypertension** | DRUG_calcium | GENE_impact | **DISEASE_hypertension** |
| 4 | DISEASE_syndrome | DRUG_cholesterol | GENE_set | GENE_large |
| 5 | **DISEASE_diabetes mellitus** | DRUG_nitric oxide | **GENE_tnf** | DISEASE_syndrome |
| 6 | **DISEASE_atrophy** | DRUG_zinc | **GENE_lep** | GENE_insulin |
| 7 | **DISEASE_insulin resistance** | DRUG_potassium | GENE_insulin receptor | **DISEASE_diabetes mellitus** |
| 8 | DISEASE_death | **DRUG_Dopamine** | GENE_renin | **DISEASE_atrophy** |
| 9 | **DISEASE_inflammation** | DRUG_aldosterone | GENE_myoglobin | **DISEASE_insulin resistance** |
| 10 | **DISEASE_myocardial infarction** | DRUG_iron | **GENE_mmp9** | DISEASE_death |
| 11 | **DISEASE_hyperglycemia** | DRUG_norepinephrine | GENE_camp | **DISEASE_inflammation** |
| 12 | **DISEASE_infection** | DRUG_glutathione | GENE_rest | **DISEASE_myocardial infarction** |
| 13 | **DISEASE_erythema** | DRUG_sucrose | GENE_cd4 | GENE_impact |
| 14 | DISEASE_body weight | DRUG_rosiglitazone | GENE_hr | **DISEASE_hyperglycemia** |
| 15 | **DISEASE_stroke** | DRUG_guanine | GENE_plasminogen | **DISEASE_infection** |
| 16 | **DISEASE_atherosclerosis** | **DRUG_dexamethasone** | GENE_albumin | DRUG_glycerol |
| 17 | **DISEASE_necrosis** | DRUG_testosterone | GENE_ar | DRUG_arachidonic acid |
| 18 | **DISEASE_heart failure** | DRUG_adenosine | **GENE_glucagon** | **DISEASE_erythema** |
| 19 | **DISEASE_fibrosis** | DRUG_ester | GENE_rhodopsin | DISEASE_body weight |
| 20 | **DISEASE_hypertrophy** | **DRUG_epinephrine** | **GENE_mmp2** | **DISEASE_stroke** |

**Table 8. A list of genes found to be in interaction with metformin by literature**

| Interaction Found | Interacting Gene with Metformin |
|---|---|
| X | Large |
| O | Insulin |
| X | Impact |
| X | Set |
| X | Hr |
| O | ca2 |
| O | Camp |
| X | Met |
| X | Rest |
| O | Albumin |
| X | Renin |
| O | insulin receptor |
| O | Plasminogen |
| O | Myoglobin |
| X | Tg |
| O | Resistin |
| O | Ace |
| X | Pah |
| O | peroxisome proliferator-activated receptor gamma |
| O | cd4 |
| X | Rhodopsin |
| X | Ar |

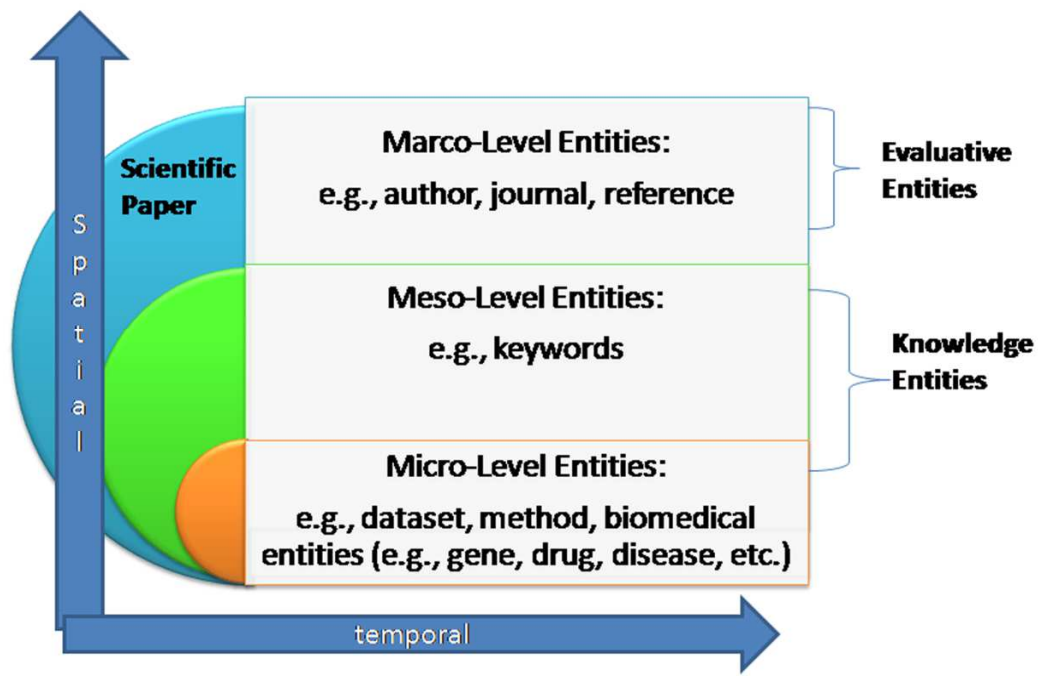

**Figure 1. Entitymetrics.**

```
ACSS1-->GENE__Q9NUB1
acyl-CoA synthetase short-chain family member 1-->GENE__Q9NUB1
ACSS2-->GENE__Q9NR19
acyl-CoA synthetase short-chain family member 2-->GENE__Q9NR19
ACSS3-->GENE__Q9H6R3
acyl-CoA synthetase short-chain family member 3-->GENE__Q9H6R3
ACTA1-->GENE__P68133
actin, alpha 1, skeletal muscle-->GENE__P68133
ACTA2-->GENE__P62736
actin, alpha 2, smooth muscle, aorta-->GENE__P62736
ACTB-->GENE__Q96HG5
actin, beta-->GENE__Q96HG5
ACTBL2-->GENE__Q562R1
actin, beta-like 2-->GENE__Q562R1
ACTC1-->GENE__P68032
actin, alpha, cardiac muscle 1-->GENE__P68032
ACTG1-->GENE__P63261
actin, gamma 1-->GENE__P63261
ACTG2-->GENE__P63267
actin, gamma 2, smooth muscle, enteric-->GENE__P63267
ACTL6A-->GENE__O96019
actin-like 6A-->GENE__O96019
ACTL6B-->GENE__O94805
actin-like 6B-->GENE__O94805
```

**Figure 2. A fraction of bio-entity dictionary.**

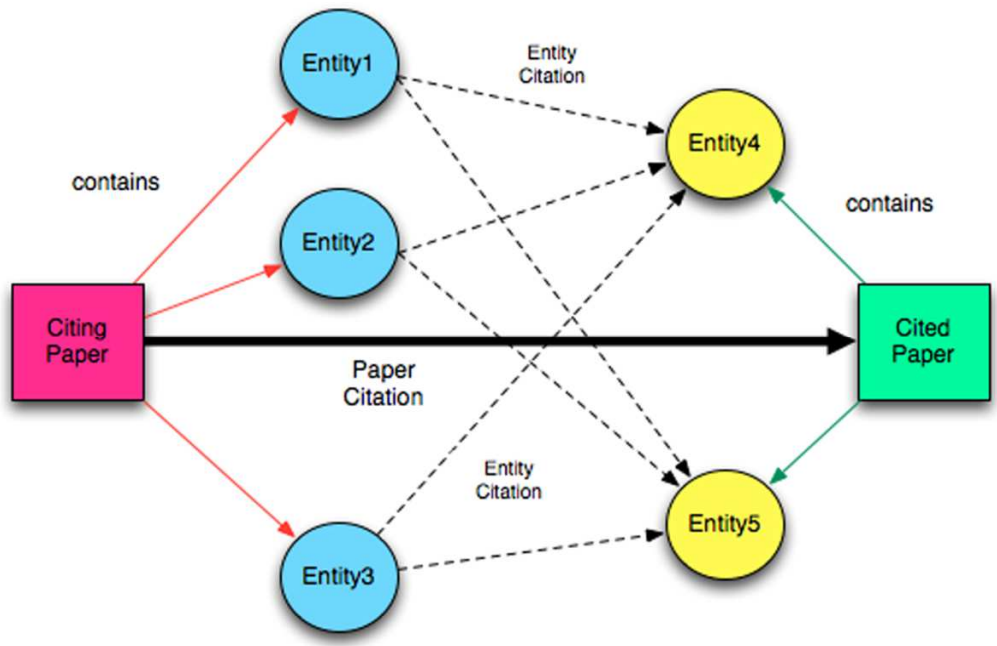

**Figure 3. From paper citation to entity citation.**

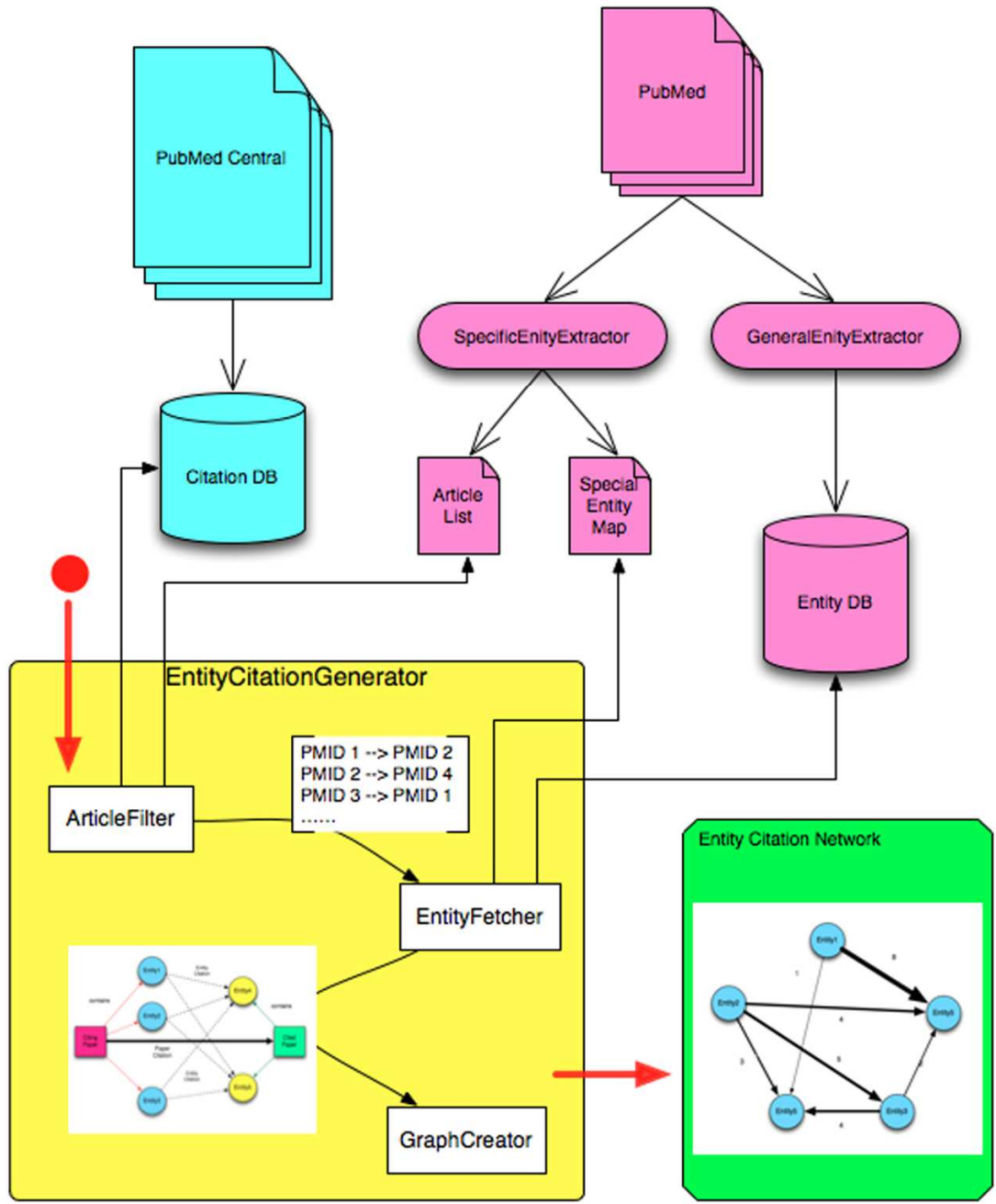

**Figure 4. An overview of Entity Citation Network generation.**

**Figure 5. The longest path of the network.**

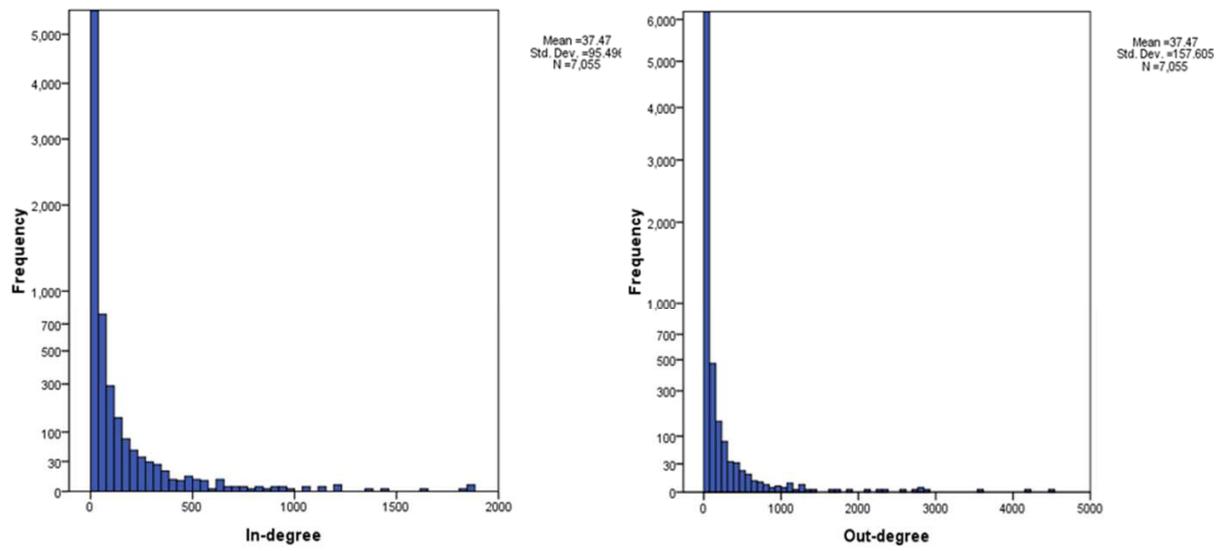

**Figure 6. In-degree and out-degree distribution of the Metformin entity-entity citation network.**